\documentclass[aps,prl,twocolumn,groupedaddress]{revtex4-1}

\usepackage{amsfonts}
\usepackage{graphicx}

\begin{document}


\title{Reflection, transmutation, annihilation and resonance in two-component kink collisions}

\author{A. Alonso-Izquierdo}
\email[]{alonsoiz@usal.es}
\affiliation{Departamento de Matematica Aplicada, University of Salamanca, Spain}

\date{\today}

\begin{abstract}
In this paper the study of collisions between kinks arising in the family of MSTB models is addressed. Phenomena such as elastic kink reflection, mutual annihilation, kink-antikink transmutation and inelastic reflection are found and depend on the impact velocity.
\end{abstract}

\pacs{11.10.Lm, 11.27.+d}

\maketitle

\section{Introduction}

Over the last decades, solitary wave solutions in nonlinear field theories have played an essential role in the explanation of new phenomena in diverse branches of Physics, e.g., Condensed Matter \cite{Davydov1985,Eschenfelder1981,Jona1993,Salje1993,Strukov2012}, Cosmology \cite{Vilenkin1994}, Optics \cite{Agrawall1995}, etc. This fact has drawn attention to the scattering of these objects. Studies on this issue concerning the kink solutions which appear in (1+1) relativistic one-component scalar field theories with a potential with two o more degenerate minima have revealed unexpected behaviors. For example, the dynamics of interacting kinks and antikinks in the archetypal $\phi^4$ model, described by Campbell, Schonfeld and Wingate in the seminal paper \cite{Campbell1983} exhibits a fascinating structure. For kink-antikink collisions where the initial relative velocity $v$ is greater than the critical speed $v_c\approx 0.2598$, these single solutions collide, bounce back and escape but if $v<v_c$ they are compelled to collide a second time. In this last case, a kink-antikink bound state (bion) is formed except for certain initial velocity ranges (resonant windows) where the kink and the antikink escape after a finite number of impacts due to the resonant energy transfer mechanism. In addition, the resulting separation velocity versus collision velocity graph displays a fractal structure \cite{Anninos1991}. An analytical explanation of this feature using the collective coordinate method is given in \cite{Goodman2008}. Similar results have been found for kink-antikink interactions in the modified sine-Gordon model \cite{Peyrard1983}, polynomial models \cite{Weigel2014, Gani2014, Dorey2011,Belendryasova2017,Saadatmand2012}, non-polynomial models \cite{Bazeia2017a,Bazeia2017b} and coupled two-component $\phi^4$ models \cite{Halavanau2012, Alonso2017}, kink-impurity interactions in the sine-Gordon and $\phi^4$ models \cite{Fei1992a,Fei1992b,Malomed1985,Malomed1989,Malomed1992}, soliton-defect interactions in the sine-Gordon model \cite{Goodman2004} and the collision of vector solitons in the coupled nonlinear Schr\"odinger model \cite{Tan2001,Yang2000}. Negative radiation pressure, where a kink hit by a plane wave is accelerated towards the source of radiation is another remarkable phenomenon which can occur in this type of models \cite{Romanczukiewicz2008,Romanczukiewicz2017}.

In this paper a new pattern in the dependence of the kink separation velocity as a function of the collision velocity is described. It arises in the one-parameter family of (1+1)-relativistic two scalar field MSTB models (named after Montonen-Sarker-Trullinger-Bishop). In this case, the potential is the fourth-degree polynomial isotropic in quartic but anisotropic in quadratic terms, $U(\phi_1,\phi_2)=\frac{1}{2}(\phi_1^2+\phi_2^2-1)^2 + \frac{1}{2}\sigma^2 \phi_2^2$ . This model is a natural generalization of the $\phi^4$ model in two-component scalar field theories, which further preserves the presence of two minima. Indeed, the MSTB model is a physical system with a proud history. In 1976 Montonen, searching for charged solitons in a model with one complex and one real scalar field, discovered by fixing the time-dependent phase for the complex field, the previously mentioned model \cite{Montonen1976}. Two different types of static topological kinks were found for the parameter range $\sigma\in (0,1)$: the first one joins the potential minima by means of a straight line, whereas the second type follows an elliptic trajectory. In a previous paper \cite{Rajaraman1975} Rajaraman and Weinberg had identified the first class of these solutions and had described the qualitative behavior of the second type in a more general model. Sarker, Trullinger and Bishop established from an energetic point of view that kink solutions of the second type are stable while those of the first type are unstable \cite{Trullinger1976}. Further analysis of kink stability in this model were performed in \cite{Currie1979,Trullinger1979}.  In 1979  Rajaraman \cite{Rajaraman1979} discovered a non-topological kink for the parameter value $\sigma=\frac{1}{2}$ whose orbit is a circle. The discovery of this new type of solitary wave prompted several numerical investigations by Subbaswamy and Trullinger. These authors numerically found that there exists a continuous family of non-topological kinks that describe closed orbits \cite{Subbaswamy1980,Subbaswamy1981}. In 1984, Magyari and Thomas \cite{Magyari1984} showed that the system of static field equations is completely integrable by finding two constants of motion. Indeed the system is not only completely integrable but Hamilton-Jacobi separable by using elliptic coordinates. This fact was used by Ito to analytically describe the whole static kink variety \cite{Ito1985}. It was proved by applying the Morse index theorem to the kink orbit manifold that the non-topological kinks are unstable \cite{Ito1985b, Guilarte1987,Guilarte1988}. In 1998 new two-component scalar field theory models that exhibit the same properties than the MSTB system were identified \cite{Alonso1998}. In 2008, a systematic classification of these generalized MSTB models was established in \cite{Alonso2008}. The extension of the MSTB model to $N$-component scalar field theories as well as the identification of the static kink manifold and the analysis of kink stability is completed in \cite{Alonso2000,Alonso2002}. Furthermore, the promotion of the MSTB model to the quantum realm is dealt with in \cite{Alonso2002b}, where the semiclassical mass of the stable static topological kinks is computed in the generalized zeta function regularization context.

The issue addressed in this paper is the study of the collisions between two stable topological kinks in the MSTB model that carry opposite topological charges although they do not form an antikink-kink pair because they describe different orbits. In this case a complex dependence of the scattering outcome with respect to the collision velocity is found: ranges of collision velocities where the kinks elastically and inelastically reflect, mutually annihilate or transmute in its antikinks coexist. In addition, sequence of resonant windows arise for some values of the model parameter $\sigma$ where the kinks collide several times before escaping and moving away.

The organization of this paper is as follows: in Section 1 the MSTB model is introduced and its static kink variety is determined; in Section 2 the scattering between stable kinks with opposite topological charges is numerically analyzed and the results are described and, finally, in Section 3 some conclusions are drawn.

\section{Model and static kinks}

We shall deal with a one-parameter family of (1+1)-dimensional two-coupled scalar field theory models whose dynamics is governed by the action
\begin{equation}
S=\int d^2x \left[ \frac{1}{2}\partial_\mu \phi_a \partial^\mu \phi_a - U[\phi_1,\phi_2] \right] \hspace{0.3cm} . \label{action}
\end{equation}
Here $\phi^a : \mathbb{R}^{1,1} \rightarrow \mathbb{R}$, $a=1,2$, are dimensionless real scalar fields and Minkowski metric $g_{\mu\nu}$ is chosen as $g_{00}=-g_{11}=1$ and $g_{12}=g_{21}=0$. The notation $x^0\equiv t$ and $x^1\equiv x$ is used from now on. The MSTB potential function $U$ in (\ref{action}) is given by
\begin{equation}
U(\phi_1,\phi_2)= \frac{1}{2} (\phi_1^2+\phi_2^2-1)^2 + \frac{\sigma^2}{2} \phi_2^2 \hspace{0.3cm}, \label{potential}
\end{equation}
where the parameter $\sigma \in(0,1)$.

\begin{figure}[h]
\includegraphics[height=3cm]{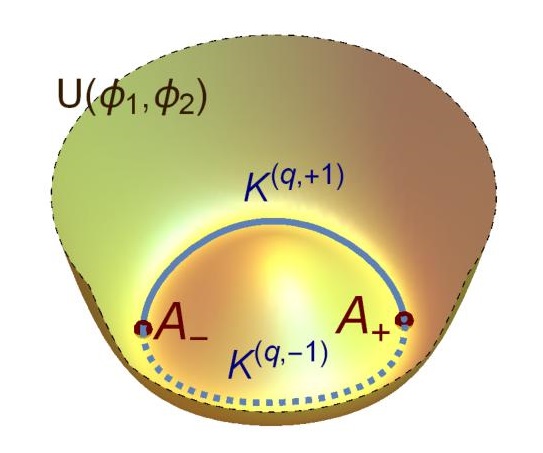}
\caption{\small MSTB potential and elliptic orbits of the stable topological kinks $K^{(q,1)}$ (solid curve) and $K^{(q,-1)}$ (dashed curve).}
\end{figure}

The Euler-Lagrange equations derived from the action (\ref{action}) lead to the coupled nonlinear Klein-Gordon equations
\begin{eqnarray}
\frac{\partial^2 \phi_1}{\partial t^2} - \frac{\partial^2 \phi_1}{\partial x^2} &=& 2 \phi_1(1-\phi_1^2-\phi_2^2)  \hspace{0.6cm} , \label{klein1}\\
\frac{\partial^2 \phi_2}{\partial t^2} - \frac{\partial^2 \phi_2}{\partial x^2} &=& 2 \phi_2 (1-\phi_1^2-\phi_2^2-{\textstyle\frac{\sigma^2}{2}}) \hspace{0.3cm}. \label{klein2}
\end{eqnarray}
The action functional (\ref{action}) is invariant by the symmetry group $\mathbb{G}=\mathbb{Z}_2\times \mathbb{Z}_2$ generated by the transformations $\pi_i:\phi_i \rightarrow -\phi_i$ with $i=1,2$.  The potential (\ref{potential}) has two degenerate absolute minima $A_\pm =(\pm 1,0)$, see Fig. 1. The solutions belonging to the finite energy configuration space ${\cal C}= \{ K(x,t)\equiv (\phi_1(x,t),\phi_2(x,t)) \in \mathbb{R}\times \mathbb{R} : E[K(x,t)] <+\infty\}$ must asymptotically connect elements of the set ${\cal M}=\{A_+,A_-\}$. This allows us to define the topological charge $q = {\textstyle\frac{1}{2}} \,  \left| \phi_1(+\infty,t) - \phi_1(-\infty,t ) \right|$, which is a physical system invariant.

The static kink variety in this model, which consists of two types of topological kinks and a family of non-topological kinks, are given as follows:

\vspace{0.1cm}

\noindent -- (1a) The four stable topological kinks
\begin{equation}
K_{\rm static}^{(q,\lambda)}(x)= \left(q \, \tanh (\sigma \overline{x}) , \lambda \sqrt{1-\sigma^2} \, {\rm sech} (\sigma \overline{x}) \right) \hspace{0.3cm}, \label{kink2}
\end{equation}
where $\overline{x}=x-x_0$ with $x_0\in \mathbb{R}$, are placed on the elliptic orbit $\phi_1^2+ \phi_2^2/(1-\sigma^2)=1$. Here $q=\pm 1$ is the topological charge and $\lambda=\pm 1$ distinguishes if the second field $\phi_2$ is positive o negative, see Fig. 1. Charge conjugation turns a kink into its antikink, i.e., $\overline{K}{}^{(q,\lambda)}(x)=K^{(q,\lambda)}(-x) = K^{(-q,\lambda)}(x)$. The energy of these solutions is $E[K_{\rm static}^{(q,\lambda)}(x)]= 2\sigma(1-\sigma^2/3)$.

\vspace{0.1cm}

\noindent -- (1b) The pair of unstable kinks
\begin{equation}
{\cal K}_{\rm static}^{(q)}(x)=(q \tanh \overline{x},0) \hspace{0.3cm}, \label{kink1}
\end{equation}
where $q=\pm 1$, connect the minima $A_+$ and $A_-$ by means of the straight line $\phi_2=0$. These solutions are more energetic than the previous kinks, $E[{\cal K}_{\rm static}^{(q)}(x)]= 4/3$.

\vspace{0.1cm}

\noindent -- (2) Finally, there exits a family of unstable non-topological kinks $N_{\rm static}(x;\gamma)=(\phi_1(x,\gamma),\phi_2(x,\gamma))$ with
\begin{eqnarray*}
\phi_1(x;\gamma)&=&\frac{(\sigma-1)(1+ e_1)+
(\sigma+1)(e_2^2+e_3)}{(\sigma-1)(1+e_1)- (\sigma+1)(e_2^2+e_3)} \hspace{0.2cm},  \nonumber \\
\phi_2(x;\gamma)&=&\frac{2(\sigma^2-1)
e_2(e_3-1)}{(\sigma-1)(1+e_1)- (\sigma+1)(e_2^2+e_3)} \hspace{0.2cm},
\label{eq:mstbntk}
\end{eqnarray*}
being $e_1=\exp[2(1+\sigma)(\bar{x}+\sigma \gamma)]$, $e_2=\exp[\sigma(\bar{x}+\gamma)]$, $e_3=\exp[2(\bar{x}+\gamma \sigma^2)]$ and $\gamma\in \mathbb{R}$. They describe closed orbits which begin and end at the point $A_+$. Similar solutions starting and ending at the point $A_-$ can be constructed by the transformation $\pi_1$. The energy sum rule $E[N(x;\gamma)]=E[K^{(q,\lambda)}(x)]+E[{\cal K}^{(q)}(x)]$ holds.

\section{Numerical analysis of the $K^{(q,\lambda)}$-$K^{(-q,-\lambda)}$ scattering}

In this Section the study of the scattering between the kinks $K^{(q,1)}$ and $K^{(-q,-1)}$ is addressed. Although these solutions carry opposite topological charge they do not form a kink-antikink pair because its trajectories are different: the $K^{(q,1)}$ kink is defined in the semiplane $\phi_2>0$ while the $K^{(-q,-1)}$ solution lives in $\phi_2<0$, see Fig. 1. The initial configuration consists of two well separated boosted static kinks
\begin{equation}
K^{(q,\lambda)}(x-x_0,t;v_0) \cup K^{(-q,-\lambda)}(x+x_0,t;-v_0) \label{conca}
\end{equation}
which are pushed together with collision velocity $v_0$. Here $K^{(q,\lambda)}(x,t;v_0)=K_{\rm static}^{(q,\lambda)} [(x-v_0 t)/\sqrt{1-v_0^2} ]$. The concatenation (\ref{conca}) describes a closed elliptic curve starting and ending at $A_-$. The non-linearity of the evolution equations (\ref{klein1}) and (\ref{klein2}) forces us to employ numerical simulations to describe the behavior of the scattering solutions. We use the modified algorithm described by Kassam and Trefethen in \cite{Kassam2005}, which is spectral in space and fourth order in time. We also complement the previous scheme with the use of the energy conservative second-order finite difference Strauss-Vazquez algorithm \cite{Strauss1978} implemented with Mur boundary conditions \cite{Mur1981}, which absorb the linear plane waves at the boundaries and let more control over the radiation evolution. The two previous numerical schemes provide identical results.

\begin{figure}[h]
\includegraphics[width=8cm]{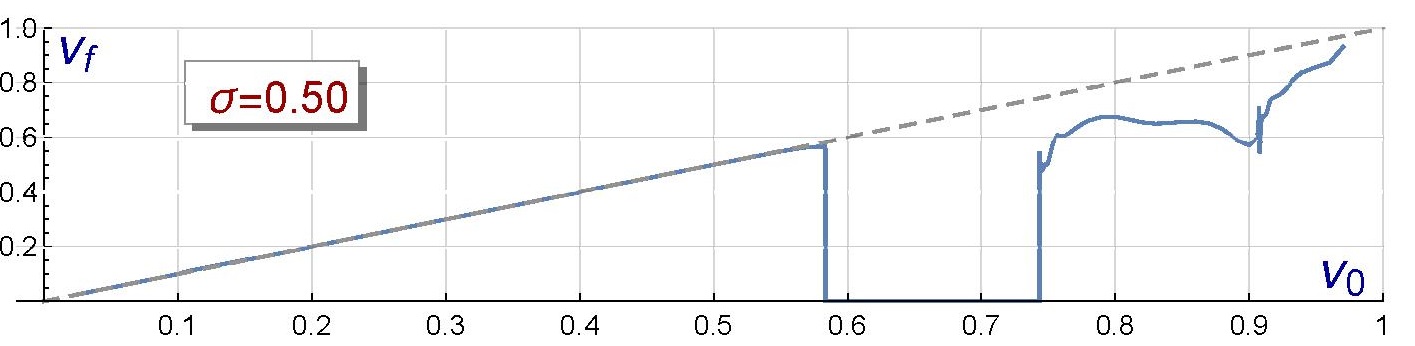} \\
\includegraphics[width=8cm]{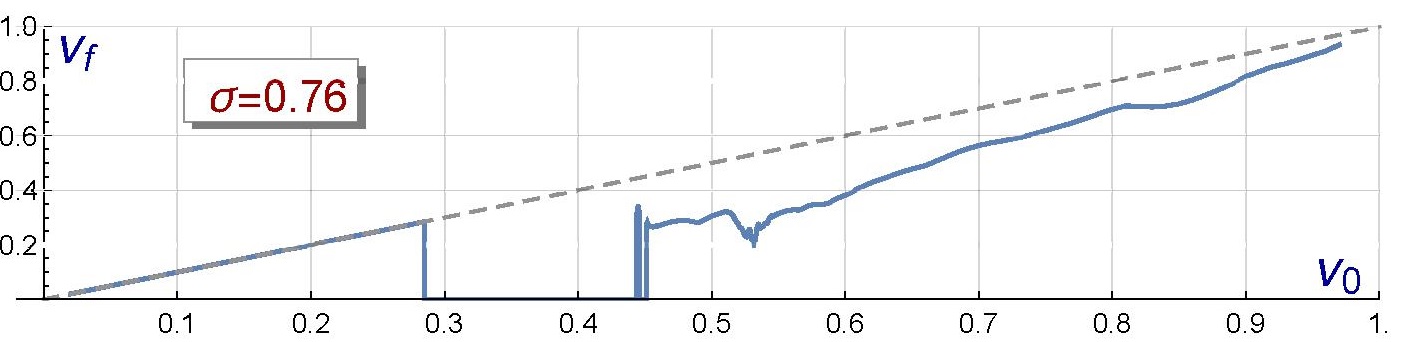} \\
\includegraphics[width=8cm]{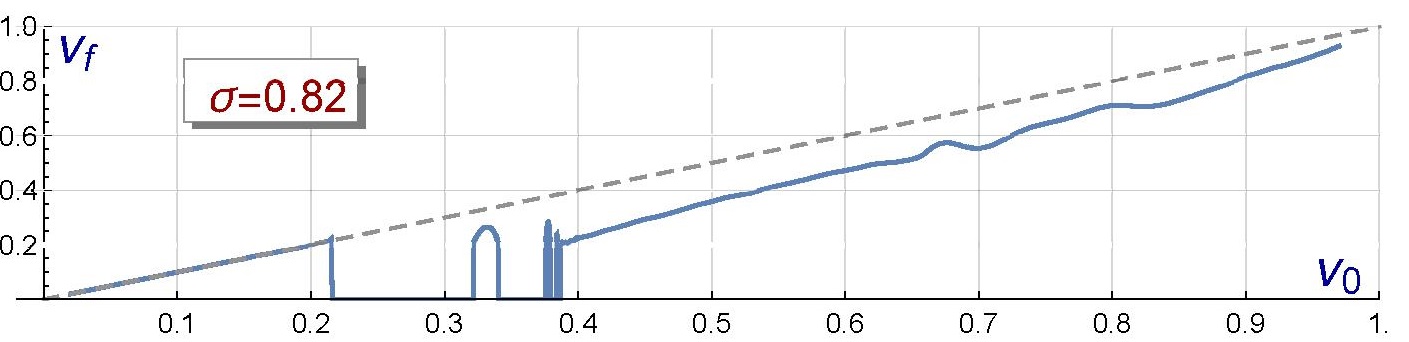}
\caption{\small Final kink velocity as a function of the initial velocity for the $K^{(q,\lambda)}$-$K^{(-q,-\lambda)}$ collisions for several values of $\sigma$. Zero final velocity indicates mutual kink annihilation. For the sake of comparison a dashed line characterizing an elastic scattering is plotted.}
\end{figure}

The dependence of the final velocity $v_f$ of the scattered kinks with respect to the impact velocity $v_0$ is displayed in Fig. 2 for several values of the parameter $\sigma$. Five types of initial velocity windows can be distinguished in these scattering processes:

\noindent -- (1) \textit{Elastic reflection windows:} For low collision velocities the kink scattering is almost elastic. This process symbolically represented as $K^{(q,\lambda)} (v_0)\cup K^{(-q,-\lambda)} (-v_0) \rightarrow K^{(q,\lambda)} (-v_0)\cup K^{(-q,-\lambda)} (v_0) $ is illustrated in Fig. 3, where the evolution of the kink components is plotted. The kink cores approach each other with initial velocity $v_0$, collide, bounce back and move away approximately with the same speed.

\begin{figure}[h]
\centerline{\includegraphics[height=2.7cm]{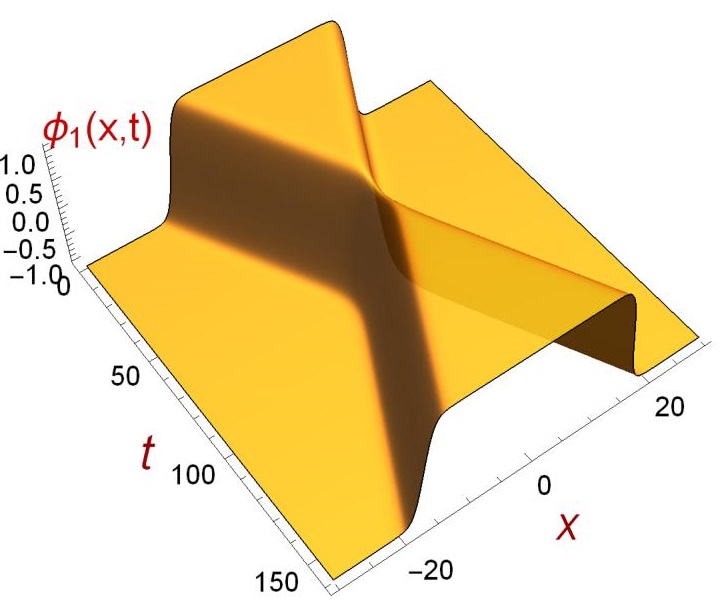} \hspace{0.3cm} \includegraphics[height=2.7cm]{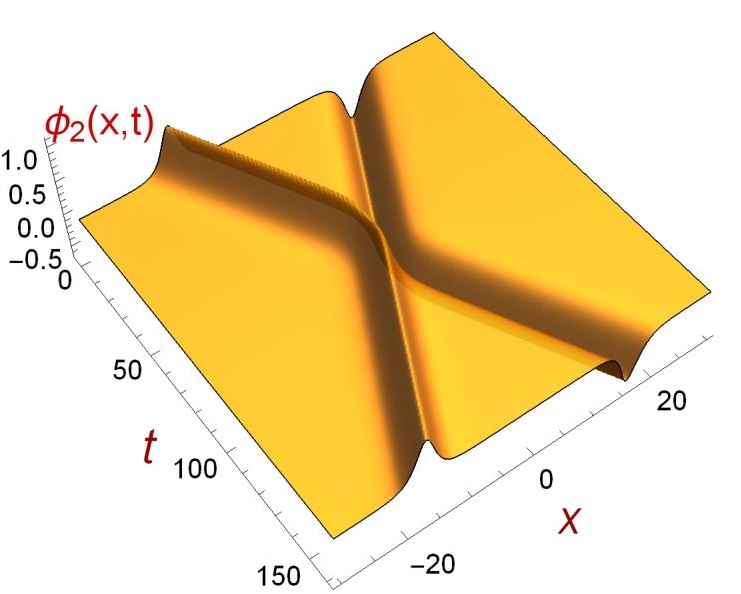}}
\caption{\small $K^{(q,\lambda)}$-$K^{(-q,-\lambda)}$ collision with impact velocity $v_0=0.25$ for the model parameter $\sigma=0.76$.}
\end{figure}

\noindent -- (2) \textit{Annihilation windows:} For these velocity intervals the kinks mutually annihilate almost instantaneously after the formation of an ephemeral bound state (bion) formed in the collision. This process characterized as $K^{(q,\lambda)} (v_0)\cup K^{(-q,-\lambda)} (-v_0) \rightarrow {\rm radiation}$ is illustrated in Fig. 4. At $t=0$ the two well separated kinks are clearly identified but after the impact the resulting configuration consists of plane waves (radiation) around the potential minimum $A_-=-1$.

\begin{figure}[h]
\centerline{\includegraphics[height=2.7cm]{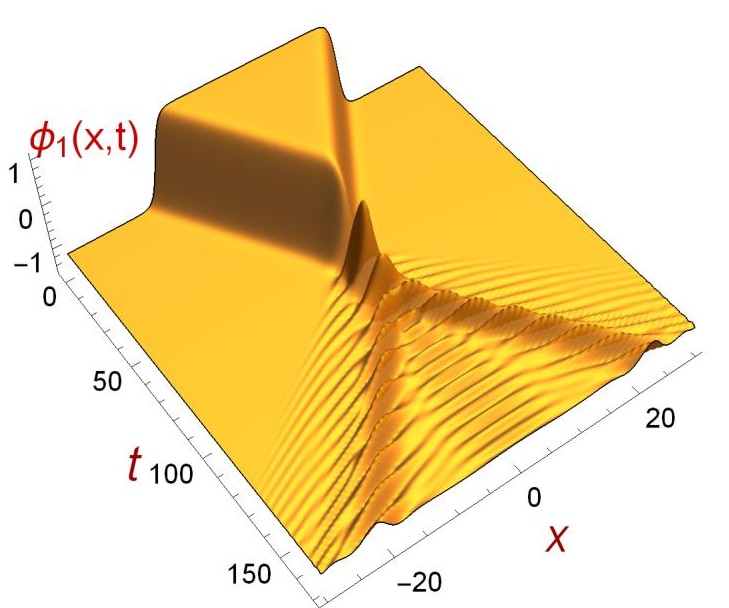} \hspace{0.3cm} \includegraphics[height=2.7cm]{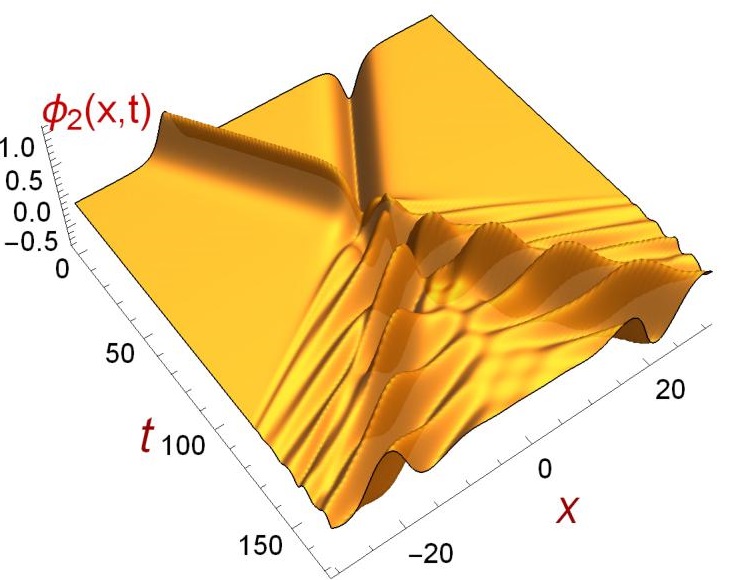}}
\caption{\small $K^{(q,\lambda)}$-$K^{(-q,-\lambda)}$ collision with impact velocity $v_0=0.4$ for the model parameter $\sigma=0.76$.}
\end{figure}

\noindent -- (3) \textit{Transmutation windows:} For some ranges of initial velocities the $K^{(q,\lambda)}$ and $K^{(-q,-\lambda)}$ kinks turn into its corresponding antikinks after the collision, see Fig. 5. This process involves the excitation of internal modes (which will be denoted by means of the asterisk superscript) and radiation emission. Therefore, we represent this event as $K^{(q,\lambda)} (v_0)\cup K^{(-q,-\lambda)} (-v_0) \rightarrow \overline{K}{}^{*(-q,-\lambda)} (-v_1)\cup \overline{K}{}^{*(q,\lambda)}  (v_1) +{\rm radiation}$ with $v_1<v_0$.

\begin{figure}[h]
\centerline{\includegraphics[height=2.7cm]{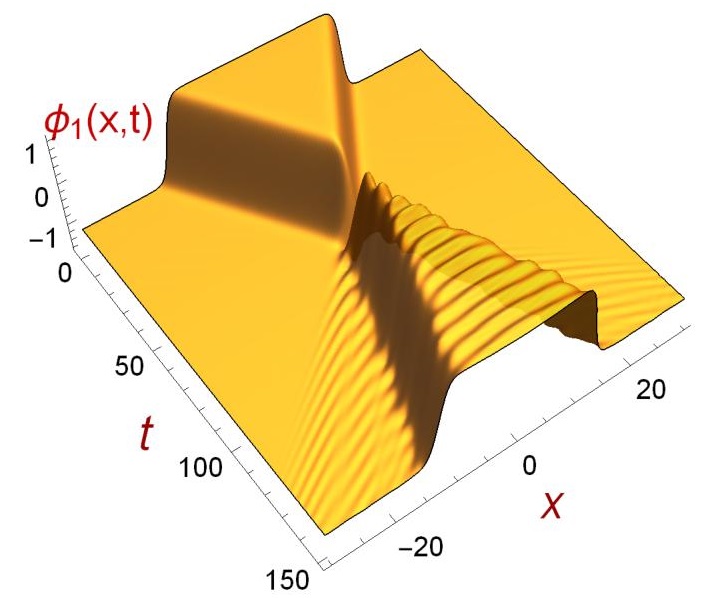} \hspace{0.3cm} \includegraphics[height=2.7cm]{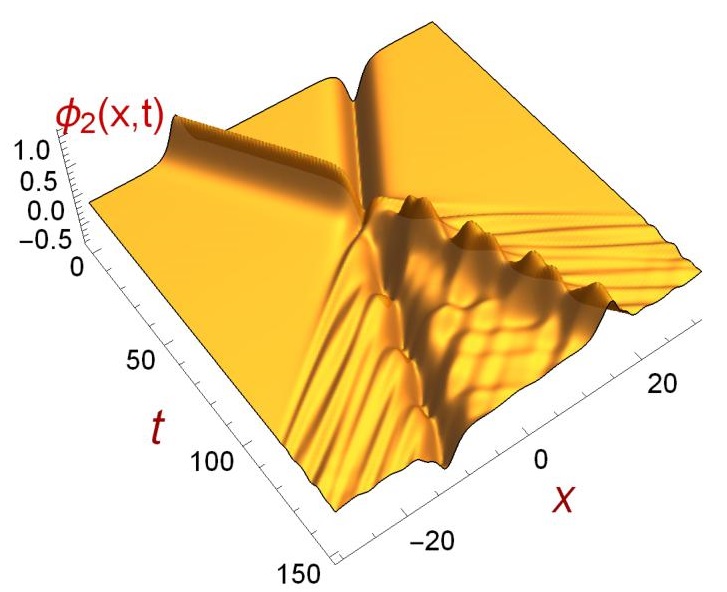}}
\caption{\small $K^{(q,\lambda)}$-$K^{(-q,-\lambda)}$ collision with impact velocity $v_0=0.5$ for the model parameter $\sigma=0.76$.}
\end{figure}

\noindent -- (4) \textit{Inelastic reflection windows:} For large enough impact velocities kink reflection occurs again but now in a non-elastic way, such that the event $K^{(q,\lambda)} (v_0)\cup K^{(-q,-\lambda)} (-v_0) \rightarrow K^{*(q,\lambda)} (-v_1)\cup K^{*(-q,-\lambda)} (v_1) + {\rm radiation}$ with $v_1<v_0$ takes place.


\noindent -- (5) \textit{Resonant windows:} For some ranges of $v_0$ a resonant energy transfer mechanism is triggered, which implies that the kinks collide and bounce back a finite number of times before recovering the kinetic energy necessary to escape. This phenomenon is illustrated in Fig. 6. Sequences of resonant windows similar to those found in the $\phi^4$ model appear for some ranges of the parameter $\sigma$, see \cite{Campbell1983}.

\begin{figure}[h]
\centerline{\includegraphics[height=2.7cm]{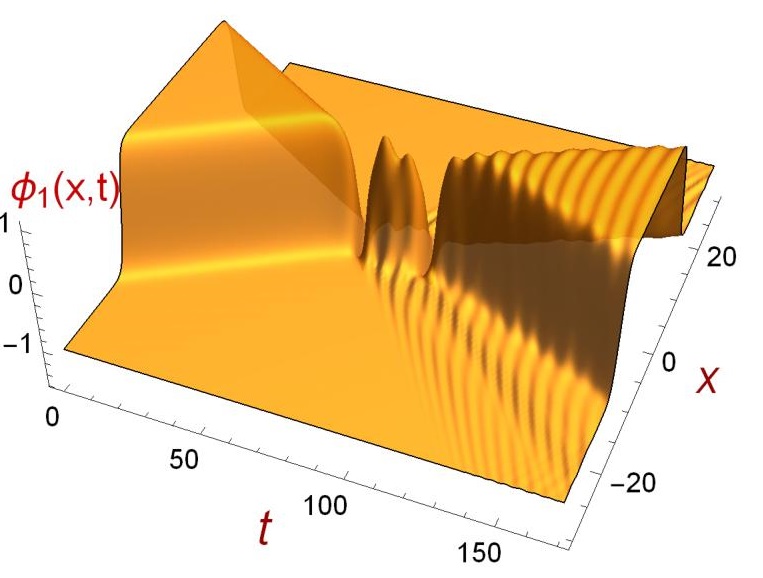} \hspace{0.3cm} \includegraphics[height=2.7cm]{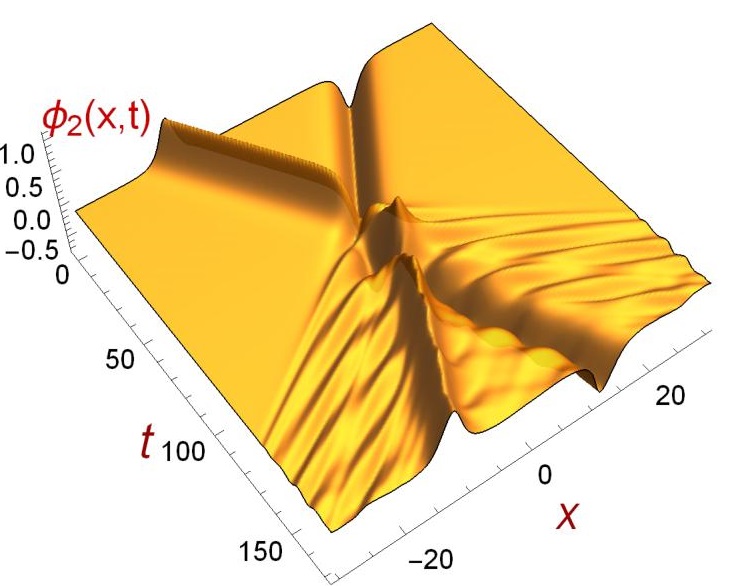}}
\caption{\small $K^{(q,\lambda)}$-$K^{(-q,-\lambda)}$ collision with impact velocity $v_0=0.445$ for the model parameter $\sigma=0.76$.}
\end{figure}

The previously described events are present in the case $\sigma=0.76$, which we have used as a benchmark in the figures introduced in this paper. In this case the elastic reflection regime is defined on the interval $(0,0.2849]$; two annihilation windows have been identified on the intervals $[0.2849,0.4434]$ and $[0.4461,0.4512]$, which confine a 2-bounce resonant window $(0.4434,0.4461)$. A quasiresonance arise at the value $v_0\approx 0.5315$, which delimitates the transmutation window $[0.4512,0.5315)$ and the inelastic reflection window $(0.5315,1)$, see Fig. 2. For $\sigma=0.5$ the resonant windows are absent while $\sigma=0.82$ involves the resonant windows $[0.3217,0.3401]$, $[0.3756,0.379]$ and $[0.3839,0.3843]$, but lacks the transmutation window.

A global vision of the behavior of the $K^{(q,\lambda)}$-$K^{(-q,-\lambda)}$ scattering in the MSTB model can be grasped from Fig. 7, where the dependence of the final velocity $v_f$ of the scattered kinks with respect to the initial kink velocity $v_0$ and the model parameter $\sigma$ can be visualized in a 3D graphic. The elastic regime for low velocities $v_0$ is clearly observed, this regime is more prevalent as the parameter $\sigma$ decreases. As the collision velocities $v_0$ increases annihilation windows appear for all the values of parameter $\sigma$. We can visualize these windows as a large cannon in Fig. 7. The presence of quasiresonances carves the valley of the landscape displayed in Fig. 7, which meets the cannon for the value $\sigma \approx 0.78$. The region delimited by the annihilation windows and the quasiresonance curve determines the transmutation windows. The inelastic reflection regime arises for impact speeds greater than the quasiresonance values and the annihilation velocity windows. Sequences of resonant windows with decreasing width emerge inside the cannon for values greater than $\sigma \approx 0.78$, which are difficult to see in the 3D graphics.

An heuristic explanation of the previously described pattern underlies the orbit evolution of the combined kink $K^{(q,\lambda)}$-$K^{(-q,-\lambda)}$. This configuration traces an elliptic orbit that starts and ends at one of the points $A_\pm$ and surrounds the local maxima exhibited by the MSTB potential at the origin in the internal space, see Fig. 1. The kink collision disturbs this loop by introducing perturbations along the $\phi_1$ and $\phi_2$ component. For low impact velocities the collision provokes small perturbations that does not change this configuration, giving rise to the elastic reflection regime. However, for initial velocities in the annihilation window the impact provokes $\phi_1$-perturbations which make the loop jump the potential maximum, the energy losses in form of radiation emission and internal mode excitations prevent the solution from returning to the loop configuration and consequently kink annihilation takes place. When the collision velocity $v_0$ is large enough the $K^{(q,\lambda)}$-$K^{(-q,-\lambda)}$ solution carries enough energy to overcome the previous situation, returning to the loop configuration. If $v_0$ belongs to the transmutation windows the induced $\phi_2$-fluctuations flip the elliptic orbit branches with positive and negative $\phi_2$, which implies the conversion of kinks into antikinks. In this sense the quasiresonances appear when the $\phi_2$-perturbations change the $K^{(q,\lambda)}$-$K^{(-q,-\lambda)}$ solution into the metastable ${\cal K}^{(q)}$-${\cal K}^{(-q)}$ configuration after the previous flip. For velocities in the inelastic reflection windows a double flip between the elliptic branches is carried out, which implies a kink reflection as final result. Finally, for certain intervals of $\sigma$ and $v_0$ a resonant energy transfer mechanism takes place where the kinks collide and bounce back $N$-times before reflecting ($N$ even) or transmuting into its antikinks ($N$ odd).

\begin{figure}[h]
\includegraphics[width=5cm]{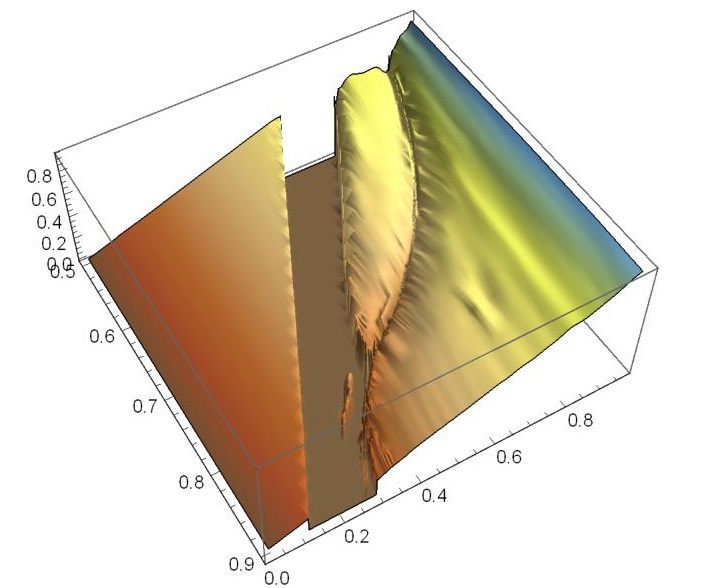}
\caption{\small Final kink velocity as a function of the initial velocity and the model parameter $\sigma$ for the $K^{(q,\lambda)}$-$K^{(-q,-\lambda)}$ collisions. Zero final velocity indicates mutual kink annihilation.}
\end{figure}

\section{Conclusions}

Our investigation of the $K^{(q,\lambda)}$-$K^{(-q,-\lambda)}$ scattering in the MSTB model has unveiled a very complex and rich variety of behaviors including elastic kink reflection, mutual annihilation, kink-antikink transmutation and inelastic reflection, whose presence depends on the impact velocity and the model parameter. A heuristic explanation based on the orbit evolution has been suggested. It remains a mayor challenge, deserving further study, to find a detailed analytical explanation based on collective coordinates or other similar techniques.



%



\begin{acknowledgments}
The author acknowledges the Spanish Ministerio de Econom\'{\i}a y Competitividad for financial support under grant MTM2014-57129-C2-1-P. They are also grateful to the Junta de Castilla y Le\'on for financial help under grant VA057U16.
\end{acknowledgments}



\begin{thebibliography}{51}%
\makeatletter
\providecommand \@ifxundefined [1]{%
 \@ifx{#1\undefined}
}%
\providecommand \@ifnum [1]{%
 \ifnum #1\expandafter \@firstoftwo
 \else \expandafter \@secondoftwo
 \fi
}%
\providecommand \@ifx [1]{%
 \ifx #1\expandafter \@firstoftwo
 \else \expandafter \@secondoftwo
 \fi
}%
\providecommand \natexlab [1]{#1}%
\providecommand \enquote  [1]{``#1''}%
\providecommand \bibnamefont  [1]{#1}%
\providecommand \bibfnamefont [1]{#1}%
\providecommand \citenamefont [1]{#1}%
\providecommand \href@noop [0]{\@secondoftwo}%
\providecommand \href [0]{\begingroup \@sanitize@url \@href}%
\providecommand \@href[1]{\@@startlink{#1}\@@href}%
\providecommand \@@href[1]{\endgroup#1\@@endlink}%
\providecommand \@sanitize@url [0]{\catcode `\\12\catcode `\$12\catcode
  `\&12\catcode `\#12\catcode `\^12\catcode `\_12\catcode `\%12\relax}%
\providecommand \@@startlink[1]{}%
\providecommand \@@endlink[0]{}%
\providecommand \url  [0]{\begingroup\@sanitize@url \@url }%
\providecommand \@url [1]{\endgroup\@href {#1}{\urlprefix }}%
\providecommand \urlprefix  [0]{URL }%
\providecommand \Eprint [0]{\href }%
\providecommand \doibase [0]{http://dx.doi.org/}%
\providecommand \selectlanguage [0]{\@gobble}%
\providecommand \bibinfo  [0]{\@secondoftwo}%
\providecommand \bibfield  [0]{\@secondoftwo}%
\providecommand \translation [1]{[#1]}%
\providecommand \BibitemOpen [0]{}%
\providecommand \bibitemStop [0]{}%
\providecommand \bibitemNoStop [0]{.\EOS\space}%
\providecommand \EOS [0]{\spacefactor3000\relax}%
\providecommand \BibitemShut  [1]{\csname bibitem#1\endcsname}%
\let\auto@bib@innerbib\@empty
\bibitem [{\citenamefont {Davydov}\ and\ \citenamefont
  {Reidel}()}]{Davydov1985}%
  \BibitemOpen
  \bibfield  {author} {\bibinfo {author} {\bibfnamefont {A.}~\bibnamefont
  {Davydov}}\ and\ \bibinfo {author} {\bibfnamefont {D.}~\bibnamefont
  {Reidel}},\ }\href@noop {} {\emph {\bibinfo {title} {Solitons in molecular
  systems,(1985) Dordrech, D}}}\ (\bibinfo  {publisher} {Reidel})\BibitemShut
  {NoStop}%
\bibitem [{\citenamefont {Eschenfelder}(2012)}]{Eschenfelder1981}%
  \BibitemOpen
  \bibfield  {author} {\bibinfo {author} {\bibfnamefont {A.~H.}\ \bibnamefont
  {Eschenfelder}},\ }\href@noop {} {\emph {\bibinfo {title} {Magnetic bubble
  technology}}},\ Vol.~\bibinfo {volume} {14}\ (\bibinfo  {publisher} {Springer
  Science \& Business Media},\ \bibinfo {year} {2012})\BibitemShut {NoStop}%
\bibitem [{\citenamefont {Jona}\ and\ \citenamefont
  {Shirane}(1993)}]{Jona1993}%
  \BibitemOpen
  \bibfield  {author} {\bibinfo {author} {\bibfnamefont {F.}~\bibnamefont
  {Jona}}\ and\ \bibinfo {author} {\bibfnamefont {G.}~\bibnamefont {Shirane}},\
  }\href@noop {} {\emph {\bibinfo {title} {Ferroelectric Crystals}}}\ (\bibinfo
   {publisher} {New York, Dover},\ \bibinfo {year} {1993})\BibitemShut
  {NoStop}%
\bibitem [{\citenamefont {Salje}(1993)}]{Salje1993}%
  \BibitemOpen
  \bibfield  {author} {\bibinfo {author} {\bibfnamefont {E.}~\bibnamefont
  {Salje}},\ }\href@noop {} {\emph {\bibinfo {title} {Phase Transitions in
  Ferroelastic and Co-Elastic Crystals}}}\ (\bibinfo  {publisher} {Cambridge
  University Press, Cambridge, UK.},\ \bibinfo {year} {1993})\BibitemShut
  {NoStop}%
\bibitem [{\citenamefont {Strukov}\ and\ \citenamefont
  {Levanyuk}(2012)}]{Strukov2012}%
  \BibitemOpen
  \bibfield  {author} {\bibinfo {author} {\bibfnamefont {B.~A.}\ \bibnamefont
  {Strukov}}\ and\ \bibinfo {author} {\bibfnamefont {A.~P.}\ \bibnamefont
  {Levanyuk}},\ }\href@noop {} {\emph {\bibinfo {title} {Ferroelectric
  phenomena in crystals: physical foundations}}}\ (\bibinfo  {publisher}
  {Springer Science \& Business Media},\ \bibinfo {year} {2012})\BibitemShut
  {NoStop}%
\bibitem [{\citenamefont {Vilenkin}\ and\ \citenamefont
  {Shellard}(2000)}]{Vilenkin1994}%
  \BibitemOpen
  \bibfield  {author} {\bibinfo {author} {\bibfnamefont {A.}~\bibnamefont
  {Vilenkin}}\ and\ \bibinfo {author} {\bibfnamefont {E.~P.~S.}\ \bibnamefont
  {Shellard}},\ }\href@noop {} {\emph {\bibinfo {title} {Cosmic strings and
  other topological defects}}}\ (\bibinfo  {publisher} {Cambridge University
  Press},\ \bibinfo {year} {2000})\BibitemShut {NoStop}%
\bibitem [{\citenamefont {Agrawall}(1995)}]{Agrawall1995}%
  \BibitemOpen
  \bibfield  {author} {\bibinfo {author} {\bibfnamefont {G.}~\bibnamefont
  {Agrawall}},\ }\href@noop {} {\emph {\bibinfo {title} {Nonlinear Fiber
  Optics}}}\ (\bibinfo  {publisher} {Academic Press},\ \bibinfo {year}
  {1995})\BibitemShut {NoStop}%
\bibitem [{\citenamefont {Campbell}\ \emph {et~al.}(1983)\citenamefont
  {Campbell}, \citenamefont {Schonfeld},\ and\ \citenamefont
  {Wingate}}]{Campbell1983}%
  \BibitemOpen
  \bibfield  {author} {\bibinfo {author} {\bibfnamefont {D.~K.}\ \bibnamefont
  {Campbell}}, \bibinfo {author} {\bibfnamefont {J.~F.}\ \bibnamefont
  {Schonfeld}}, \ and\ \bibinfo {author} {\bibfnamefont {C.~A.}\ \bibnamefont
  {Wingate}},\ }\href {\doibase http://dx.doi.org/10.1016/0167-2789(83)90289-0}
  {\bibfield  {journal} {\bibinfo  {journal} {Physica D: Nonlinear Phenomena}\
  }\textbf {\bibinfo {volume} {9}},\ \bibinfo {pages} {1} (\bibinfo {year}
  {1983})}\BibitemShut {NoStop}%
\bibitem [{\citenamefont {Anninos}\ \emph {et~al.}(1991)\citenamefont
  {Anninos}, \citenamefont {Oliveira},\ and\ \citenamefont
  {Matzner}}]{Anninos1991}%
  \BibitemOpen
  \bibfield  {author} {\bibinfo {author} {\bibfnamefont {P.}~\bibnamefont
  {Anninos}}, \bibinfo {author} {\bibfnamefont {S.}~\bibnamefont {Oliveira}}, \
  and\ \bibinfo {author} {\bibfnamefont {R.~A.}\ \bibnamefont {Matzner}},\
  }\href {\doibase 10.1103/PhysRevD.44.1147} {\bibfield  {journal} {\bibinfo
  {journal} {Phys. Rev. D}\ }\textbf {\bibinfo {volume} {44}},\ \bibinfo
  {pages} {1147} (\bibinfo {year} {1991})}\BibitemShut {NoStop}%
\bibitem [{\citenamefont {Goodman}(2008)}]{Goodman2008}%
  \BibitemOpen
  \bibfield  {author} {\bibinfo {author} {\bibfnamefont {R.~H.}\ \bibnamefont
  {Goodman}},\ }\href {\doibase 10.1063/1.2904823} {\bibfield  {journal}
  {\bibinfo  {journal} {Chaos: An Interdisciplinary Journal of Nonlinear
  Science}\ }\textbf {\bibinfo {volume} {18}},\ \bibinfo {pages} {023113}
  (\bibinfo {year} {2008})},\ \Eprint
  {http://arxiv.org/abs/http://dx.doi.org/10.1063/1.2904823}
  {http://dx.doi.org/10.1063/1.2904823} \BibitemShut {NoStop}%
\bibitem [{\citenamefont {Peyrard}\ and\ \citenamefont
  {Campbell}(1983)}]{Peyrard1983}%
  \BibitemOpen
  \bibfield  {author} {\bibinfo {author} {\bibfnamefont {M.}~\bibnamefont
  {Peyrard}}\ and\ \bibinfo {author} {\bibfnamefont {D.~K.}\ \bibnamefont
  {Campbell}},\ }\href {\doibase
  http://dx.doi.org/10.1016/0167-2789(83)90290-7} {\bibfield  {journal}
  {\bibinfo  {journal} {Physica D: Nonlinear Phenomena}\ }\textbf {\bibinfo
  {volume} {9}},\ \bibinfo {pages} {33 } (\bibinfo {year} {1983})}\BibitemShut
  {NoStop}%
\bibitem [{\citenamefont {Weigel}(2014)}]{Weigel2014}%
  \BibitemOpen
  \bibfield  {author} {\bibinfo {author} {\bibfnamefont {H.}~\bibnamefont
  {Weigel}},\ }\href@noop {} {\bibfield  {journal} {\bibinfo  {journal}
  {Journal of Physics: Conference Series}\ }\textbf {\bibinfo {volume} {482}},\
  \bibinfo {pages} {012045} (\bibinfo {year} {2014})}\BibitemShut {NoStop}%
\bibitem [{\citenamefont {Gani}\ \emph {et~al.}(2014)\citenamefont {Gani},
  \citenamefont {Kudryavtsev},\ and\ \citenamefont {Lizunova}}]{Gani2014}%
  \BibitemOpen
  \bibfield  {author} {\bibinfo {author} {\bibfnamefont {V.~A.}\ \bibnamefont
  {Gani}}, \bibinfo {author} {\bibfnamefont {A.~E.}\ \bibnamefont
  {Kudryavtsev}}, \ and\ \bibinfo {author} {\bibfnamefont {M.~A.}\ \bibnamefont
  {Lizunova}},\ }\href {\doibase 10.1103/PhysRevD.89.125009} {\bibfield
  {journal} {\bibinfo  {journal} {Phys. Rev. D}\ }\textbf {\bibinfo {volume}
  {89}},\ \bibinfo {pages} {125009} (\bibinfo {year} {2014})}\BibitemShut
  {NoStop}%
\bibitem [{\citenamefont {Dorey}\ \emph {et~al.}(2011)\citenamefont {Dorey},
  \citenamefont {Mersh}, \citenamefont {Romanczukiewicz},\ and\ \citenamefont
  {Shnir}}]{Dorey2011}%
  \BibitemOpen
  \bibfield  {author} {\bibinfo {author} {\bibfnamefont {P.}~\bibnamefont
  {Dorey}}, \bibinfo {author} {\bibfnamefont {K.}~\bibnamefont {Mersh}},
  \bibinfo {author} {\bibfnamefont {T.}~\bibnamefont {Romanczukiewicz}}, \ and\
  \bibinfo {author} {\bibfnamefont {Y.}~\bibnamefont {Shnir}},\ }\href
  {\doibase 10.1103/PhysRevLett.107.091602} {\bibfield  {journal} {\bibinfo
  {journal} {Phys. Rev. Lett.}\ }\textbf {\bibinfo {volume} {107}},\ \bibinfo
  {pages} {091602} (\bibinfo {year} {2011})}\BibitemShut {NoStop}%
\bibitem [{\citenamefont {Belendryasova}\ and\ \citenamefont
  {Gani}(2017)}]{Belendryasova2017}%
  \BibitemOpen
  \bibfield  {author} {\bibinfo {author} {\bibfnamefont {E.}~\bibnamefont
  {Belendryasova}}\ and\ \bibinfo {author} {\bibfnamefont {V.}~\bibnamefont
  {Gani}},\ }\href@noop {} {\enquote {\bibinfo {title} {Scattering of the
  $\phi^8$ kinks with power-law asymptotics},}\ } (\bibinfo {year} {2017}),\
  \Eprint {http://arxiv.org/abs/hep-th/170800403} {arXiv:hep-th/170800403}
  \BibitemShut {NoStop}%
\bibitem [{\citenamefont {Saadatmand}\ and\ \citenamefont
  {Javidan}(2012)}]{Saadatmand2012}%
  \BibitemOpen
  \bibfield  {author} {\bibinfo {author} {\bibfnamefont {D.}~\bibnamefont
  {Saadatmand}}\ and\ \bibinfo {author} {\bibfnamefont {K.}~\bibnamefont
  {Javidan}},\ }\href {http://stacks.iop.org/1402-4896/85/i=2/a=025003}
  {\bibfield  {journal} {\bibinfo  {journal} {Physica Scripta}\ }\textbf
  {\bibinfo {volume} {85}},\ \bibinfo {pages} {025003} (\bibinfo {year}
  {2012})}\BibitemShut {NoStop}%
\bibitem [{\citenamefont {Bazeia}\ \emph
  {et~al.}(2017{\natexlab{a}})\citenamefont {Bazeia}, \citenamefont
  {Belendryasova},\ and\ \citenamefont {Gani}}]{Bazeia2017a}%
  \BibitemOpen
  \bibfield  {author} {\bibinfo {author} {\bibfnamefont {D.}~\bibnamefont
  {Bazeia}}, \bibinfo {author} {\bibfnamefont {E.}~\bibnamefont
  {Belendryasova}}, \ and\ \bibinfo {author} {\bibfnamefont {V.}~\bibnamefont
  {Gani}},\ }\href@noop {} {\enquote {\bibinfo {title} {Scattering of kinks in
  a non-polynomial model},}\ } (\bibinfo {year} {2017}{\natexlab{a}}),\ \Eprint
  {http://arxiv.org/abs/hep-th/171107788} {arXiv:hep-th/171107788} \BibitemShut
  {NoStop}%
\bibitem [{\citenamefont {Bazeia}\ \emph
  {et~al.}(2017{\natexlab{b}})\citenamefont {Bazeia}, \citenamefont
  {Belendryasova},\ and\ \citenamefont {Gani}}]{Bazeia2017b}%
  \BibitemOpen
  \bibfield  {author} {\bibinfo {author} {\bibfnamefont {D.}~\bibnamefont
  {Bazeia}}, \bibinfo {author} {\bibfnamefont {E.}~\bibnamefont
  {Belendryasova}}, \ and\ \bibinfo {author} {\bibfnamefont {V.}~\bibnamefont
  {Gani}},\ }\href@noop {} {\enquote {\bibinfo {title} {Scattering of kinks of
  the sinh-deformed $\phi^4$ model},}\ } (\bibinfo {year}
  {2017}{\natexlab{b}}),\ \Eprint {http://arxiv.org/abs/hep-th/171004993}
  {arXiv:hep-th/171004993} \BibitemShut {NoStop}%
\bibitem [{\citenamefont {Halavanau}\ \emph {et~al.}(2012)\citenamefont
  {Halavanau}, \citenamefont {Romanczukiewicz},\ and\ \citenamefont
  {Shnir}}]{Halavanau2012}%
  \BibitemOpen
  \bibfield  {author} {\bibinfo {author} {\bibfnamefont {A.}~\bibnamefont
  {Halavanau}}, \bibinfo {author} {\bibfnamefont {T.}~\bibnamefont
  {Romanczukiewicz}}, \ and\ \bibinfo {author} {\bibfnamefont {Y.}~\bibnamefont
  {Shnir}},\ }\href {\doibase 10.1103/PhysRevD.86.085027} {\bibfield  {journal}
  {\bibinfo  {journal} {Phys. Rev. D}\ }\textbf {\bibinfo {volume} {86}},\
  \bibinfo {pages} {085027} (\bibinfo {year} {2012})}\BibitemShut {NoStop}%
\bibitem [{\citenamefont {Alonso-Izquierdo}(2017)}]{Alonso2017}%
  \BibitemOpen
  \bibfield  {author} {\bibinfo {author} {\bibfnamefont {A.}~\bibnamefont
  {Alonso-Izquierdo}},\ }\href {\doibase
  https://doi.org/10.1016/j.physd.2017.10.006} {\bibfield  {journal} {\bibinfo
  {journal} {Physica D: Nonlinear Phenomena}\ } (\bibinfo {year} {2017}),\
  https://doi.org/10.1016/j.physd.2017.10.006}\BibitemShut {NoStop}%
\bibitem [{\citenamefont {Fei}\ \emph {et~al.}(1992{\natexlab{a}})\citenamefont
  {Fei}, \citenamefont {Kivshar},\ and\ \citenamefont {V\'azquez}}]{Fei1992a}%
  \BibitemOpen
  \bibfield  {author} {\bibinfo {author} {\bibfnamefont {Z.}~\bibnamefont
  {Fei}}, \bibinfo {author} {\bibfnamefont {Y.~S.}\ \bibnamefont {Kivshar}}, \
  and\ \bibinfo {author} {\bibfnamefont {L.}~\bibnamefont {V\'azquez}},\ }\href
  {\doibase 10.1103/PhysRevA.45.6019} {\bibfield  {journal} {\bibinfo
  {journal} {Phys. Rev. A}\ }\textbf {\bibinfo {volume} {45}},\ \bibinfo
  {pages} {6019} (\bibinfo {year} {1992}{\natexlab{a}})}\BibitemShut {NoStop}%
\bibitem [{\citenamefont {Fei}\ \emph {et~al.}(1992{\natexlab{b}})\citenamefont
  {Fei}, \citenamefont {Kivshar},\ and\ \citenamefont {V\'azquez}}]{Fei1992b}%
  \BibitemOpen
  \bibfield  {author} {\bibinfo {author} {\bibfnamefont {Z.}~\bibnamefont
  {Fei}}, \bibinfo {author} {\bibfnamefont {Y.~S.}\ \bibnamefont {Kivshar}}, \
  and\ \bibinfo {author} {\bibfnamefont {L.}~\bibnamefont {V\'azquez}},\ }\href
  {\doibase 10.1103/PhysRevA.45.6019} {\bibfield  {journal} {\bibinfo
  {journal} {Phys. Rev. A}\ }\textbf {\bibinfo {volume} {45}},\ \bibinfo
  {pages} {6019} (\bibinfo {year} {1992}{\natexlab{b}})}\BibitemShut {NoStop}%
\bibitem [{\citenamefont {Malomed}(1985)}]{Malomed1985}%
  \BibitemOpen
  \bibfield  {author} {\bibinfo {author} {\bibfnamefont {B.~A.}\ \bibnamefont
  {Malomed}},\ }\href {\doibase https://doi.org/10.1016/S0167-2789(85)80006-3}
  {\bibfield  {journal} {\bibinfo  {journal} {Physica D: Nonlinear Phenomena}\
  }\textbf {\bibinfo {volume} {15}},\ \bibinfo {pages} {385 } (\bibinfo {year}
  {1985})}\BibitemShut {NoStop}%
\bibitem [{\citenamefont {Malomed}(1989)}]{Malomed1989}%
  \BibitemOpen
  \bibfield  {author} {\bibinfo {author} {\bibfnamefont {B.~A.}\ \bibnamefont
  {Malomed}},\ }\href {\doibase https://doi.org/10.1016/0375-9601(89)90422-2}
  {\bibfield  {journal} {\bibinfo  {journal} {Physics Letters A}\ }\textbf
  {\bibinfo {volume} {136}},\ \bibinfo {pages} {395 } (\bibinfo {year}
  {1989})}\BibitemShut {NoStop}%
\bibitem [{\citenamefont {Malomed}(1992)}]{Malomed1992}%
  \BibitemOpen
  \bibfield  {author} {\bibinfo {author} {\bibfnamefont {B.~A.}\ \bibnamefont
  {Malomed}},\ }\href {http://stacks.iop.org/0305-4470/25/i=4/a=015} {\bibfield
   {journal} {\bibinfo  {journal} {Journal of Physics A: Mathematical and
  General}\ }\textbf {\bibinfo {volume} {25}},\ \bibinfo {pages} {755}
  (\bibinfo {year} {1992})}\BibitemShut {NoStop}%
\bibitem [{\citenamefont {Goodman}\ and\ \citenamefont
  {Haberman}(2004)}]{Goodman2004}%
  \BibitemOpen
  \bibfield  {author} {\bibinfo {author} {\bibfnamefont {R.~H.}\ \bibnamefont
  {Goodman}}\ and\ \bibinfo {author} {\bibfnamefont {R.}~\bibnamefont
  {Haberman}},\ }\href {\doibase http://dx.doi.org/10.1016/j.physd.2004.04.002}
  {\bibfield  {journal} {\bibinfo  {journal} {Physica D: Nonlinear Phenomena}\
  }\textbf {\bibinfo {volume} {195}},\ \bibinfo {pages} {303 } (\bibinfo {year}
  {2004})}\BibitemShut {NoStop}%
\bibitem [{\citenamefont {Tan}\ and\ \citenamefont {Yang}(2001)}]{Tan2001}%
  \BibitemOpen
  \bibfield  {author} {\bibinfo {author} {\bibfnamefont {Y.}~\bibnamefont
  {Tan}}\ and\ \bibinfo {author} {\bibfnamefont {J.}~\bibnamefont {Yang}},\
  }\href {\doibase 10.1103/PhysRevE.64.056616} {\bibfield  {journal} {\bibinfo
  {journal} {Phys. Rev. E}\ }\textbf {\bibinfo {volume} {64}},\ \bibinfo
  {pages} {056616} (\bibinfo {year} {2001})}\BibitemShut {NoStop}%
\bibitem [{\citenamefont {Yang}\ and\ \citenamefont {Tan}(2000)}]{Yang2000}%
  \BibitemOpen
  \bibfield  {author} {\bibinfo {author} {\bibfnamefont {J.}~\bibnamefont
  {Yang}}\ and\ \bibinfo {author} {\bibfnamefont {Y.}~\bibnamefont {Tan}},\
  }\href {\doibase 10.1103/PhysRevLett.85.3624} {\bibfield  {journal} {\bibinfo
   {journal} {Phys. Rev. Lett.}\ }\textbf {\bibinfo {volume} {85}},\ \bibinfo
  {pages} {3624} (\bibinfo {year} {2000})}\BibitemShut {NoStop}%
\bibitem [{\citenamefont {Romanczukiewicz}(2008)}]{Romanczukiewicz2008}%
  \BibitemOpen
  \bibfield  {author} {\bibinfo {author} {\bibfnamefont {T.}~\bibnamefont
  {Romanczukiewicz}},\ }\href@noop {} {\bibfield  {journal} {\bibinfo
  {journal} {Acta Phys. Polon. B}\ }\textbf {\bibinfo {volume} {29}},\ \bibinfo
  {pages} {3449} (\bibinfo {year} {2008})}\BibitemShut {NoStop}%
\bibitem [{\citenamefont {Romanczukiewicz}(2017)}]{Romanczukiewicz2017}%
  \BibitemOpen
  \bibfield  {author} {\bibinfo {author} {\bibfnamefont {T.}~\bibnamefont
  {Romanczukiewicz}},\ }\href {\doibase
  https://doi.org/10.1016/j.physletb.2017.08.045} {\bibfield  {journal}
  {\bibinfo  {journal} {Physics Letters B}\ }\textbf {\bibinfo {volume}
  {773}},\ \bibinfo {pages} {295 } (\bibinfo {year} {2017})}\BibitemShut
  {NoStop}%
\bibitem [{\citenamefont {Montonen}(1976)}]{Montonen1976}%
  \BibitemOpen
  \bibfield  {author} {\bibinfo {author} {\bibfnamefont {C.}~\bibnamefont
  {Montonen}},\ }\href {\doibase https://doi.org/10.1016/0550-3213(76)90537-X}
  {\bibfield  {journal} {\bibinfo  {journal} {Nuclear Physics B}\ }\textbf
  {\bibinfo {volume} {112}},\ \bibinfo {pages} {349 } (\bibinfo {year}
  {1976})}\BibitemShut {NoStop}%
\bibitem [{\citenamefont {Rajaraman}\ and\ \citenamefont
  {Weinberg}(1975)}]{Rajaraman1975}%
  \BibitemOpen
  \bibfield  {author} {\bibinfo {author} {\bibfnamefont {R.}~\bibnamefont
  {Rajaraman}}\ and\ \bibinfo {author} {\bibfnamefont {E.~J.}\ \bibnamefont
  {Weinberg}},\ }\href {\doibase 10.1103/PhysRevD.11.2950} {\bibfield
  {journal} {\bibinfo  {journal} {Phys. Rev. D}\ }\textbf {\bibinfo {volume}
  {11}},\ \bibinfo {pages} {2950} (\bibinfo {year} {1975})}\BibitemShut
  {NoStop}%
\bibitem [{\citenamefont {Sarker}\ \emph {et~al.}(1976)\citenamefont {Sarker},
  \citenamefont {Trullinger},\ and\ \citenamefont {Bishop}}]{Trullinger1976}%
  \BibitemOpen
  \bibfield  {author} {\bibinfo {author} {\bibfnamefont {S.}~\bibnamefont
  {Sarker}}, \bibinfo {author} {\bibfnamefont {S.}~\bibnamefont {Trullinger}},
  \ and\ \bibinfo {author} {\bibfnamefont {A.}~\bibnamefont {Bishop}},\ }\href
  {\doibase https://doi.org/10.1016/0375-9601(76)90784-2} {\bibfield  {journal}
  {\bibinfo  {journal} {Physics Letters A}\ }\textbf {\bibinfo {volume} {59}},\
  \bibinfo {pages} {255 } (\bibinfo {year} {1976})}\BibitemShut {NoStop}%
\bibitem [{\citenamefont {Currie}\ \emph {et~al.}(1979)\citenamefont {Currie},
  \citenamefont {Sarker}, \citenamefont {Bishop},\ and\ \citenamefont
  {Trullinger}}]{Currie1979}%
  \BibitemOpen
  \bibfield  {author} {\bibinfo {author} {\bibfnamefont {J.~F.}\ \bibnamefont
  {Currie}}, \bibinfo {author} {\bibfnamefont {S.}~\bibnamefont {Sarker}},
  \bibinfo {author} {\bibfnamefont {A.~R.}\ \bibnamefont {Bishop}}, \ and\
  \bibinfo {author} {\bibfnamefont {S.~E.}\ \bibnamefont {Trullinger}},\ }\href
  {\doibase 10.1103/PhysRevA.20.2213} {\bibfield  {journal} {\bibinfo
  {journal} {Phys. Rev. A}\ }\textbf {\bibinfo {volume} {20}},\ \bibinfo
  {pages} {2213} (\bibinfo {year} {1979})}\BibitemShut {NoStop}%
\bibitem [{\citenamefont {Trullinger}\ and\ \citenamefont
  {DeLeonardis}(1980)}]{Trullinger1979}%
  \BibitemOpen
  \bibfield  {author} {\bibinfo {author} {\bibfnamefont {S.~E.}\ \bibnamefont
  {Trullinger}}\ and\ \bibinfo {author} {\bibfnamefont {R.~M.}\ \bibnamefont
  {DeLeonardis}},\ }\href {\doibase 10.1103/PhysRevB.22.5522} {\bibfield
  {journal} {\bibinfo  {journal} {Phys. Rev. B}\ }\textbf {\bibinfo {volume}
  {22}},\ \bibinfo {pages} {5522} (\bibinfo {year} {1980})}\BibitemShut
  {NoStop}%
\bibitem [{\citenamefont {Rajaraman}(1979)}]{Rajaraman1979}%
  \BibitemOpen
  \bibfield  {author} {\bibinfo {author} {\bibfnamefont {R.}~\bibnamefont
  {Rajaraman}},\ }\href {\doibase 10.1103/PhysRevLett.42.200} {\bibfield
  {journal} {\bibinfo  {journal} {Phys. Rev. Lett.}\ }\textbf {\bibinfo
  {volume} {42}},\ \bibinfo {pages} {200} (\bibinfo {year} {1979})}\BibitemShut
  {NoStop}%
\bibitem [{\citenamefont {Subbaswamy}\ and\ \citenamefont
  {Trullinger}(1980)}]{Subbaswamy1980}%
  \BibitemOpen
  \bibfield  {author} {\bibinfo {author} {\bibfnamefont {K.~R.}\ \bibnamefont
  {Subbaswamy}}\ and\ \bibinfo {author} {\bibfnamefont {S.~E.}\ \bibnamefont
  {Trullinger}},\ }\href {\doibase 10.1103/PhysRevD.22.1495} {\bibfield
  {journal} {\bibinfo  {journal} {Phys. Rev. D}\ }\textbf {\bibinfo {volume}
  {22}},\ \bibinfo {pages} {1495} (\bibinfo {year} {1980})}\BibitemShut
  {NoStop}%
\bibitem [{\citenamefont {Subbaswamy}\ and\ \citenamefont
  {Trullinger}(1981)}]{Subbaswamy1981}%
  \BibitemOpen
  \bibfield  {author} {\bibinfo {author} {\bibfnamefont {K.}~\bibnamefont
  {Subbaswamy}}\ and\ \bibinfo {author} {\bibfnamefont {S.}~\bibnamefont
  {Trullinger}},\ }\href {\doibase
  https://doi.org/10.1016/0167-2789(81)90016-6} {\bibfield  {journal} {\bibinfo
   {journal} {Physica D: Nonlinear Phenomena}\ }\textbf {\bibinfo {volume}
  {2}},\ \bibinfo {pages} {379 } (\bibinfo {year} {1981})}\BibitemShut
  {NoStop}%
\bibitem [{\citenamefont {Magyari}\ and\ \citenamefont
  {Thomas}(1984)}]{Magyari1984}%
  \BibitemOpen
  \bibfield  {author} {\bibinfo {author} {\bibfnamefont {E.}~\bibnamefont
  {Magyari}}\ and\ \bibinfo {author} {\bibfnamefont {H.}~\bibnamefont
  {Thomas}},\ }\href {\doibase https://doi.org/10.1016/0375-9601(84)90342-6}
  {\bibfield  {journal} {\bibinfo  {journal} {Physics Letters A}\ }\textbf
  {\bibinfo {volume} {100}},\ \bibinfo {pages} {11 } (\bibinfo {year}
  {1984})}\BibitemShut {NoStop}%
\bibitem [{\citenamefont {Ito}(1985)}]{Ito1985}%
  \BibitemOpen
  \bibfield  {author} {\bibinfo {author} {\bibfnamefont {H.}~\bibnamefont
  {Ito}},\ }\href {\doibase https://doi.org/10.1016/0375-9601(85)90670-X}
  {\bibfield  {journal} {\bibinfo  {journal} {Physics Letters A}\ }\textbf
  {\bibinfo {volume} {112}},\ \bibinfo {pages} {119} (\bibinfo {year}
  {1985})}\BibitemShut {NoStop}%
\bibitem [{\citenamefont {Ito}\ and\ \citenamefont {Tasaki}(1985)}]{Ito1985b}%
  \BibitemOpen
  \bibfield  {author} {\bibinfo {author} {\bibfnamefont {H.}~\bibnamefont
  {Ito}}\ and\ \bibinfo {author} {\bibfnamefont {H.}~\bibnamefont {Tasaki}},\
  }\href {\doibase https://doi.org/10.1016/0375-9601(85)90134-3} {\bibfield
  {journal} {\bibinfo  {journal} {Physics Letters A}\ }\textbf {\bibinfo
  {volume} {113}},\ \bibinfo {pages} {179 } (\bibinfo {year}
  {1985})}\BibitemShut {NoStop}%
\bibitem [{\citenamefont {Guilarte}(1987)}]{Guilarte1987}%
  \BibitemOpen
  \bibfield  {author} {\bibinfo {author} {\bibfnamefont {J.~M.}\ \bibnamefont
  {Guilarte}},\ }\href {\doibase 10.1007/BF00420308} {\bibfield  {journal}
  {\bibinfo  {journal} {Lett. Math. Phys.}\ }\textbf {\bibinfo {volume} {14}},\
  \bibinfo {pages} {169} (\bibinfo {year} {1987})}\BibitemShut {NoStop}%
\bibitem [{\citenamefont {Guilarte}(1988)}]{Guilarte1988}%
  \BibitemOpen
  \bibfield  {author} {\bibinfo {author} {\bibfnamefont {J.~M.}\ \bibnamefont
  {Guilarte}},\ }\href {\doibase https://doi.org/10.1016/0003-4916(88)90104-2}
  {\bibfield  {journal} {\bibinfo  {journal} {Annals of Physics}\ }\textbf
  {\bibinfo {volume} {188}},\ \bibinfo {pages} {307 } (\bibinfo {year}
  {1988})}\BibitemShut {NoStop}%
\bibitem [{\citenamefont {Alonso-Izquierdo}\ \emph {et~al.}(1998)\citenamefont
  {Alonso-Izquierdo}, \citenamefont {Leon},\ and\ \citenamefont
  {Guilarte}}]{Alonso1998}%
  \BibitemOpen
  \bibfield  {author} {\bibinfo {author} {\bibfnamefont {A.}~\bibnamefont
  {Alonso-Izquierdo}}, \bibinfo {author} {\bibfnamefont {M.~A.~G.}\
  \bibnamefont {Leon}}, \ and\ \bibinfo {author} {\bibfnamefont {J.~M.}\
  \bibnamefont {Guilarte}},\ }\href
  {http://stacks.iop.org/0305-4470/31/i=1/a=021} {\bibfield  {journal}
  {\bibinfo  {journal} {Journal of Physics A: Mathematical and General}\
  }\textbf {\bibinfo {volume} {31}},\ \bibinfo {pages} {209} (\bibinfo {year}
  {1998})}\BibitemShut {NoStop}%
\bibitem [{\citenamefont {Alonso-Izquierdo}\ and\ \citenamefont
  {Guilarte}(2008)}]{Alonso2008}%
  \BibitemOpen
  \bibfield  {author} {\bibinfo {author} {\bibfnamefont {A.}~\bibnamefont
  {Alonso-Izquierdo}}\ and\ \bibinfo {author} {\bibfnamefont {J.~M.}\
  \bibnamefont {Guilarte}},\ }\href {\doibase
  https://doi.org/10.1016/j.physd.2008.07.020} {\bibfield  {journal} {\bibinfo
  {journal} {Physica D: Nonlinear Phenomena}\ }\textbf {\bibinfo {volume}
  {237}},\ \bibinfo {pages} {3263 } (\bibinfo {year} {2008})}\BibitemShut
  {NoStop}%
\bibitem [{\citenamefont {Alonso-Izquierdo}\ \emph {et~al.}(2000)\citenamefont
  {Alonso-Izquierdo}, \citenamefont {Le\'on},\ and\ \citenamefont
  {Guilarte}}]{Alonso2000}%
  \BibitemOpen
  \bibfield  {author} {\bibinfo {author} {\bibfnamefont {A.}~\bibnamefont
  {Alonso-Izquierdo}}, \bibinfo {author} {\bibfnamefont {M.~A.~G.}\
  \bibnamefont {Le\'on}}, \ and\ \bibinfo {author} {\bibfnamefont {J.~M.}\
  \bibnamefont {Guilarte}},\ }\href
  {http://stacks.iop.org/0951-7715/13/i=4/a=309} {\bibfield  {journal}
  {\bibinfo  {journal} {Nonlinearity}\ }\textbf {\bibinfo {volume} {13}},\
  \bibinfo {pages} {1137} (\bibinfo {year} {2000})}\BibitemShut {NoStop}%
\bibitem [{\citenamefont {Alonso-Izquierdo}\ \emph
  {et~al.}(2002{\natexlab{a}})\citenamefont {Alonso-Izquierdo}, \citenamefont
  {Le\'on},\ and\ \citenamefont {Guilarte}}]{Alonso2002}%
  \BibitemOpen
  \bibfield  {author} {\bibinfo {author} {\bibfnamefont {A.}~\bibnamefont
  {Alonso-Izquierdo}}, \bibinfo {author} {\bibfnamefont {M.~A.~G.}\
  \bibnamefont {Le\'on}}, \ and\ \bibinfo {author} {\bibfnamefont {J.~M.}\
  \bibnamefont {Guilarte}},\ }\href
  {http://stacks.iop.org/0951-7715/15/i=4/a=308} {\bibfield  {journal}
  {\bibinfo  {journal} {Nonlinearity}\ }\textbf {\bibinfo {volume} {15}},\
  \bibinfo {pages} {1097} (\bibinfo {year} {2002}{\natexlab{a}})}\BibitemShut
  {NoStop}%
\bibitem [{\citenamefont {Alonso-Izquierdo}\ \emph
  {et~al.}(2002{\natexlab{b}})\citenamefont {Alonso-Izquierdo}, \citenamefont
  {Fuertes}, \citenamefont {Le\'on},\ and\ \citenamefont
  {Guilarte}}]{Alonso2002b}%
  \BibitemOpen
  \bibfield  {author} {\bibinfo {author} {\bibfnamefont {A.}~\bibnamefont
  {Alonso-Izquierdo}}, \bibinfo {author} {\bibfnamefont {W.~G.}\ \bibnamefont
  {Fuertes}}, \bibinfo {author} {\bibfnamefont {M.~G.}\ \bibnamefont {Le\'on}},
  \ and\ \bibinfo {author} {\bibfnamefont {J.~M.}\ \bibnamefont {Guilarte}},\
  }\href {\doibase https://doi.org/10.1016/S0550-3213(02)00498-4} {\bibfield
  {journal} {\bibinfo  {journal} {Nuclear Physics B}\ }\textbf {\bibinfo
  {volume} {638}},\ \bibinfo {pages} {378 } (\bibinfo {year}
  {2002}{\natexlab{b}})}\BibitemShut {NoStop}%
\bibitem [{\citenamefont {Kassam}\ and\ \citenamefont
  {Trefethen}(2005)}]{Kassam2005}%
  \BibitemOpen
  \bibfield  {author} {\bibinfo {author} {\bibfnamefont {A.-K.}\ \bibnamefont
  {Kassam}}\ and\ \bibinfo {author} {\bibfnamefont {L.~N.}\ \bibnamefont
  {Trefethen}},\ }\href {\doibase 10.1137/S1064827502410633} {\bibfield
  {journal} {\bibinfo  {journal} {SIAM Journal on Scientific Computing}\
  }\textbf {\bibinfo {volume} {26}},\ \bibinfo {pages} {1214} (\bibinfo {year}
  {2005})},\ \Eprint
  {http://arxiv.org/abs/https://doi.org/10.1137/S1064827502410633}
  {https://doi.org/10.1137/S1064827502410633} \BibitemShut {NoStop}%
\bibitem [{\citenamefont {Strauss}\ and\ \citenamefont
  {Vazquez}(1978)}]{Strauss1978}%
  \BibitemOpen
  \bibfield  {author} {\bibinfo {author} {\bibfnamefont {W.}~\bibnamefont
  {Strauss}}\ and\ \bibinfo {author} {\bibfnamefont {L.}~\bibnamefont
  {Vazquez}},\ }\href {\doibase http://dx.doi.org/10.1016/0021-9991(78)90038-4}
  {\bibfield  {journal} {\bibinfo  {journal} {Journal of Computational
  Physics}\ }\textbf {\bibinfo {volume} {28}},\ \bibinfo {pages} {271 }
  (\bibinfo {year} {1978})}\BibitemShut {NoStop}%
\bibitem [{\citenamefont {Mur}(1981)}]{Mur1981}%
  \BibitemOpen
  \bibfield  {author} {\bibinfo {author} {\bibfnamefont {G.}~\bibnamefont
  {Mur}},\ }\href@noop {} {\bibfield  {journal} {\bibinfo  {journal} {IEEE
  Transactions on Electromagnetic Compatibility}\ }\textbf {\bibinfo {volume}
  {EMC-231}},\ \bibinfo {pages} {377 } (\bibinfo {year} {1981})}\BibitemShut
  {NoStop}%
\end{thebibliography}

%

\end{document}